\documentclass[a4paper,10pt]{article}
\usepackage[utf8]{inputenc}
\usepackage{layout}
\usepackage{geometry}
\usepackage{setspace}
\usepackage{ulem}
\usepackage{eurosym}
\usepackage{bookman}
\usepackage{newcent}
\usepackage{lmodern}
\usepackage{mathpazo}
\usepackage{mathptmx}
\usepackage{url}
\usepackage{verbatim}
\usepackage{listings}
\usepackage{fancyhdr}
\usepackage{color}
\usepackage{colortbl}
\usepackage{amsmath}
\usepackage{amssymb}
\title{}

\author{}

\begin{document}
\begin{center}
{\Large\bf Unimodular $f(T)$ gravity}\\

\medskip

S. B. Nassur$^{(a)}$\footnote{e-mail:nassurmaeva@gmail.com}, C. Ainamon$^{(a)}$\footnote{email:ainamoncyrille@yahoo.fr}, M. J. S. Houndjo$^{(a,b)}$\footnote{e-mail: sthoundjo@yahoo.fr} and J. Tossa$^{(a)}$\footnote{e-mail: joel.tossa@imsp-uac.org}

$^a$ \,{\it Institut de Math\'{e}matiques et de Sciences Physiques (IMSP)}\\
 {\it 01 BP 613,  Porto-Novo, B\'{e}nin}\\

$^{b}$\,{\it Facult\'e des Sciences et Techniques de Natitingou - Universit\'e de Natitingou - B\'enin} \\

\date{}

%\maketitle
\end{center}
\begin{abstract}
We reconstruct the geometrical $f(T)$ actions in the framework of unimodular $f(T)$ gravity.  The unimodular $f(T)$ gravity yields stunning properties related to the generalized Friedmann equations. Indeed, it has been found that depending on the form of the  Friedmann equations, the Lagrange multipliers may or not depend on the time parameter $\tau$. Moreover  we find that the reconstruction of $f(T)$ functions can be easily performed in general, not depending on a given scale factor, or can determine a particular way, depending on a given scale factor, in the vacuum. It is noted that the reconstruction of a general action joins is consistent to the unimodular gravity for the constant $\Lambda$.
\end{abstract}
%%%%%%%%%%%%%%%%%%%%%%%%%%%%%%%%%%%%%%%%%%%%%%%%%%%%%%%%%%%%%%%%%%%%%%SECTION%%%%%%%%%%%%%%%%%%%%%%%%%%%%%%%%%%%%%%%%%%%%%%%%%%%%%%%%%%%%%%%%%%%%%%
%%%%%%%%%%%%%%%%%%%%%%%%%%%%%%%%%%%%%%%%%%%%%%%%%%%%%%%%%%%%%%%%%%%%%%SECTION%%%%%%%%%%%%%%%%%%%%%%%%%%%%%%%%%%%%%%%%%%%%%%%%%%%%%%%%%%%%%%%%%%%%%%
%%%%%%%%%%%%%%%%%%%%%%%%%%%%%%%%%%%%%%%%%%%%%%%%%%%%%%%%%%%%%%%%%%%%%%SECTION%%%%%%%%%%%%%%%%%%%%%%%%%%%%%%%%%%%%%%%%%%%%%%%%%%%%%%%%%%%%%%%%%%%%%%
%%%%%%%%%%%%%%%%%%%%%%%%%%%%%%%%%%%%%%%%%%%%%%%%%%%%%%%%%%%%%%%%%%%%%%SECTION%%%%%%%%%%%%%%%%%%%%%%%%%%%%%%%%%%%%%%%%%%%%%%%%%%%%%%%%%%%%%%%%%%%%%%
%%%%%%%%%%%%%%%%%%%%%%%%%%%%%%%%%%%%%%%%%%%%%%%%%%%%%%%%%%%%%%%%%%%%%%SECTION%%%%%%%%%%%%%%%%%%%%%%%%%%%%%%%%%%%%%%%%%%%%%%%%%%%%%%%%%%%%%%%%%%%%%%
\section{Introduction}\label{sec1}
%%%%%%%%%%%%%%%%%%%%%%%%%%%%%%%%%%%%%%%%%%%%%%%%%%%%%%%%%%%%%%%%%%%%%%SECTION%%%%%%%%%%%%%%%%%%%%%%%%%%%%%%%%%%%%%%%%%%%%%%%%%%%%%%%%%%%%%%%%%%%%%%
%%%%%%%%%%%%%%%%%%%%%%%%%%%%%%%%%%%%%%%%%%%%%%%%%%%%%%%%%%%%%%%%%%%%%%SECTION%%%%%%%%%%%%%%%%%%%%%%%%%%%%%%%%%%%%%%%%%%%%%%%%%%%%%%%%%%%%%%%%%%%%%%
%%%%%%%%%%%%%%%%%%%%%%%%%%%%%%%%%%%%%%%%%%%%%%%%%%%%%%%%%%%%%%%%%%%%%%SECTION%%%%%%%%%%%%%%%%%%%%%%%%%%%%%%%%%%%%%%%%%%%%%%%%%%%%%%%%%%%%%%%%%%%%%%
%%%%%%%%%%%%%%%%%%%%%%%%%%%%%%%%%%%%%%%%%%%%%%%%%%%%%%%%%%%%%%%%%%%%%%SECTION%%%%%%%%%%%%%%%%%%%%%%%%%%%%%%%%%%%%%%%%%%%%%%%%%%%%%%%%%%%%%%%%%%%%%%
%%%%%%%%%%%%%%%%%%%%%%%%%%%%%%%%%%%%%%%%%%%%%%%%%%%%%%%%%%%%%%%%%%%%%%SECTION%%%%%%%%%%%%%%%%%%%%%%%%%%%%%%%%%%%%%%%%%%%%%%%%%%%%%%%%%%%%%%%%%%%%%%

Considered as an alternative to the theory of general relativity \cite{George} by Einstein in 1919, the unimodular gravity emerged
as a natural way to regain the cosmological constant. The cosmological constant could appear in the unimodular gravity
as an integration constant \cite{Finkelstein,George}, unlike general relativity, where the cosmological constant must be introduced by hand. 
The unimodular gravity is different from General Relativity that {\it its Lagrangian density is taken equal to the square of the curvature
scalar associated with the Levi-Civita connection $R^{2}$} and also that {\it the determinant of the metric tensor of space-time
is fixed} \cite{Abassi,Naveen}. Unlike general relativity where {\it Lagrangian density is taken as the scalar curvature of
Riemannian tensor $R$} and that {\it the determinant of the metric tensor of space-time is variable}. However despite these differences, 
these two theories seem equivalent \cite{Abassi}. For more information about the unimodular gravity see \cite{George,Pankaj, Kluson, Naveen, Carlos, Antonio, Ippocratis, Astrid, Gao, Basak, Cho, Alvarez} \par
The presence of dark energy (or cosmological constant) in the universe has boosted the birth of alternative theories of relativity
General (GR) including $f(R)$ theory which is characterized by a geometric Lagrangian density taken as a function of the curvature $R$, 
(for more information about $f(R)$ gravity see \cite{Nojiri11,Kauzar,Nojiri}).\par
Note that the unimodular gravity, general relativity and $f(R)$ gravity embraced the idea that gravity is felt in
spacetime through the Levi-Civita connection but it turns out that the consideration of the Levi-Civita connection is only an arbitrary 
choice to feel gravitaion in space-time.\par 
Indeed, gravity can also be felt in space-time through the Weizenbock connection. The use of the Weizenbock connection in spacetime to describe 
gravitation leads to the teleparallel theory which is also equivalent to general relativity to describe gravitation. 
The geometric Lagrangian density of teleparallel theory is taken as the scalar torsion $T$ associated with the Weizenbock connection\cite{Arcos}.\par 
The presence of dark energy in the universe also cause the modification of teleparallel theory. The most popular modification of teleparallel 
theory is that of considering the geometric part of action as  an algebraic function of scalar torsion so-call $f(T)$ thory \cite{Rafael,Abbas}.\par
Our goal in this paper is to reconstruct the geometric actions in the context of the unimodular $f(T)$ gravity in the flat universe 
Friedmann-Robertson-Walker (FRW). Both approaches present in unimodular $f(T)$ gravity unimodular will be highlighted.
One that allows easy determination of the action $f(T)$ without the use an explicit expression of the scale factor, in the vaccum
and an other one, it allows the reconstruction action $f(T)$ for a giving specific expression of the scalar factor.\par 
So let's begin our work by generalities in section\ref{sec2}: the subsection \ref{subsec1} will be sent to
generality of teleparallel theory that are the same as those of $f(T)$ theory; the subsection \ref{subsec2} will be devoted to
the unimodular approach considered; as to the subsection\ref{subsec3}, it will be for the methods of reconstruction of unimodular $f(T)$ gravity. Then we will begin reconstruction of the specific action $f(T)$ in unimodular $f(T)$ gravity in Section \ref{sec3}.  
Finally a conclusion is given in Section \ref{sec4}.

%%%%%%%%%%%%%%%%%%%%%%%%%%%%%%%%%%%%%%%%%%%%%%%%%%%%%%%%%%%%%%%%%%%%%%SECTION%%%%%%%%%%%%%%%%%%%%%%%%%%%%%%%%%%%%%%%%%%%%%%%%%%%%%%%%%%%%%%%%%%%%%%
%%%%%%%%%%%%%%%%%%%%%%%%%%%%%%%%%%%%%%%%%%%%%%%%%%%%%%%%%%%%%%%%%%%%%%SECTION%%%%%%%%%%%%%%%%%%%%%%%%%%%%%%%%%%%%%%%%%%%%%%%%%%%%%%%%%%%%%%%%%%%%%%
%%%%%%%%%%%%%%%%%%%%%%%%%%%%%%%%%%%%%%%%%%%%%%%%%%%%%%%%%%%%%%%%%%%%%%SECTION%%%%%%%%%%%%%%%%%%%%%%%%%%%%%%%%%%%%%%%%%%%%%%%%%%%%%%%%%%%%%%%%%%%%%%
%%%%%%%%%%%%%%%%%%%%%%%%%%%%%%%%%%%%%%%%%%%%%%%%%%%%%%%%%%%%%%%%%%%%%%SECTION%%%%%%%%%%%%%%%%%%%%%%%%%%%%%%%%%%%%%%%%%%%%%%%%%%%%%%%%%%%%%%%%%%%%%%
%%%%%%%%%%%%%%%%%%%%%%%%%%%%%%%%%%%%%%%%%%%%%%%%%%%%%%%%%%%%%%%%%%%%%%SECTION%%%%%%%%%%%%%%%%%%%%%%%%%%%%%%%%%%%%%%%%%%%%%%%%%%%%%%%%%%%%%%%%%%%%%%
\section{Generality}\label{sec2}
%%%%%%%%%%%%%%%%%%%%%%%%%%%%%%%%%%%%%%%%%%%%%%%%%%%%%%%%%%%%%%%%%%%%%%SECTION%%%%%%%%%%%%%%%%%%%%%%%%%%%%%%%%%%%%%%%%%%%%%%%%%%%%%%%%%%%%%%%%%%%%%%
%%%%%%%%%%%%%%%%%%%%%%%%%%%%%%%%%%%%%%%%%%%%%%%%%%%%%%%%%%%%%%%%%%%%%%SECTION%%%%%%%%%%%%%%%%%%%%%%%%%%%%%%%%%%%%%%%%%%%%%%%%%%%%%%%%%%%%%%%%%%%%%%
%%%%%%%%%%%%%%%%%%%%%%%%%%%%%%%%%%%%%%%%%%%%%%%%%%%%%%%%%%%%%%%%%%%%%%SECTION%%%%%%%%%%%%%%%%%%%%%%%%%%%%%%%%%%%%%%%%%%%%%%%%%%%%%%%%%%%%%%%%%%%%%%
%%%%%%%%%%%%%%%%%%%%%%%%%%%%%%%%%%%%%%%%%%%%%%%%%%%%%%%%%%%%%%%%%%%%%%%%%%%%%%%%%%%%%%%%%%%%%%%%%%%%%%%%%%%%%%%%%%%%%%%%%%%%%%%%%%%%%%%%%%%%%%%%%%%
%%%%%%%%%%%%%%%%%%%%%%%%%%%%%%%%%%%%%%%%%%%%%%%%%%%%%%%%%%%%%%%%%%%%%%%%%%%%%%%%%%%%%%%%%%%%%%%%%%%%%%%%%%%%%%%%%%%%%%%%%%%%%%%%%%%%%%%%%%%%%%%%%%%
%%%%%%%%%%%%%%%%%%%%%%%%%%%%%%%%%%%%%%%%%%%%%%%%%%%%%%%%%%%%%%%%%%%%%%%%%%%%%%%%%%%%%%%%%%%%%%%%%%%%%%%%%%%%%%%%%%%%%%%%%%%%%%%%%%%%%%%%%%%%%%%%%%%
%%%%%%%%%%%%%%%%%%%%%%%%%%%%%%%%%%%%%%%%%%%%%%%%%%%%%%%%%%%%%%%%%%%%%%SUBSECTION%%%%%%%%%%%%%%%%%%%%%%%%%%%%%%%%%%%%%%%%%%%%%%%%%%%%%%%%%%%%%%%%%%%%%%
%%%%%%%%%%%%%%%%%%%%%%%%%%%%%%%%%%%%%%%%%%%%%%%%%%%%%%%%%%%%%%%%%%%%%%SUBSECTION%%%%%%%%%%%%%%%%%%%%%%%%%%%%%%%%%%%%%%%%%%%%%%%%%%%%%%%%%%%%%%%%%%%%%%
%%%%%%%%%%%%%%%%%%%%%%%%%%%%%%%%%%%%%%%%%%%%%%%%%%%%%%%%%%%%%%%%%%%%%%SUBSECTION%%%%%%%%%%%%%%%%%%%%%%%%%%%%%%%%%%%%%%%%%%%%%%%%%%%%%%%%%%%%%%%%%%%%%%
\subsection{Preliminary $f(T)$ theory}\label{subsec1}
%%%%%%%%%%%%%%%%%%%%%%%%%%%%%%%%%%%%%%%%%%%%%%%%%%%%%%%%%%%%%%%%%%%%%%SUBSECTION%%%%%%%%%%%%%%%%%%%%%%%%%%%%%%%%%%%%%%%%%%%%%%%%%%%%%%%%%%%%%%%%%%%%%%
%%%%%%%%%%%%%%%%%%%%%%%%%%%%%%%%%%%%%%%%%%%%%%%%%%%%%%%%%%%%%%%%%%%%%%SUBSECTION%%%%%%%%%%%%%%%%%%%%%%%%%%%%%%%%%%%%%%%%%%%%%%%%%%%%%%%%%%%%%%%%%%%%%%
%%%%%%%%%%%%%%%%%%%%%%%%%%%%%%%%%%%%%%%%%%%%%%%%%%%%%%%%%%%%%%%%%%%%%%SUBSECTION%%%%%%%%%%%%%%%%%%%%%%%%%%%%%%%%%%%%%%%%%%%%%%%%%%%%%%%%%%%%%%%%%%%%%%
The modified theory of gravity based on the torsion scalar is the one for which the geometric part of the action is an algebraic function depending on the torsion. In the same way as in the Teleparallel gravity, the geometric elements are described using orthonormal tetrads components defined in the tangent space at each point of the manifold. In general the line element can be written as
\begin{eqnarray}
ds^2=g_{\mu\nu}dx^\mu dx^\nu=\eta_{ij}\theta^i\theta^j\,,
\end{eqnarray}
where we define the following elements
\begin{eqnarray}
dx^\mu=e_{i}^{\;\;\mu}\theta^{i}\,\quad \theta^{i}=e^{i}_{\;\;\mu}dx^{\mu}.
\end{eqnarray}
Note that $\eta_{ij}=diag(1,-1,-1,-1)$ is the metric related to the  Minkowskian spacetime and the $\{e^{i}_{\;\mu}\}$ are the components  of the tetrad which satisfy the following identity
\begin{eqnarray}
e^{\;\;\mu}_{i}e^{i}_{\;\;\nu}=\delta^{\mu}_{\nu},\quad e^{\;\;i}_{\mu}e^{\mu}_{\;\;j}=\delta^{i}_{j}.
\end{eqnarray}
The connection in use in this theory is the one of  Weizenbock's,  defined by
\begin{eqnarray}
\Gamma^{\lambda}_{\mu\nu}=e^{\;\;\lambda}_{i}\partial_{\mu}e^{i}_{\;\;\nu}=-e^{i}_{\;\;\mu}\partial_\nu e_{i}^{\;\;\lambda}.
\end{eqnarray}
Once the previous connection is assumed, one can then expression the main geometric objects; the torsion tensor's components as\begin{eqnarray}
T^{\lambda}_{\;\;\;\mu\nu}= \Gamma^{\lambda}_{\mu\nu}-\Gamma^{\lambda}_{\nu\mu},
\end{eqnarray}
which is used in the definition of the contorsion tensor as
\begin{eqnarray}
K^{\mu\nu}_{\;\;\;\;\lambda}=-\frac{1}{2}\left(T^{\mu\nu}_{\;\;\;\lambda}-T^{\nu\mu}_{\;\;\;\;\lambda}+T^{\;\;\;\nu\mu}_{\lambda}\right)\,\,.
\end{eqnarray}
The above objects (torsion and contorsion) are used to define a new tensor $S_{\lambda}^{\;\;\mu\nu}$ as
\begin{eqnarray}
S_{\lambda}^{\;\;\mu\nu}=\frac{1}{2}\left(K^{\mu\nu}_{\;\;\;\;\lambda}+\delta^{\mu}_{\lambda}T^{\alpha\nu}_{\;\;\;\;\alpha}-\delta^{\nu}_{\lambda}T^{\alpha\mu}_{\;\;\;\;\alpha}\right)\,\,.
\end{eqnarray}
The torsion scalar is defined from the previous tensor and the torsion tensor as
\begin{eqnarray}
T=T^{\lambda}_{\;\;\;\mu\nu}S^{\;\;\;\mu\nu}_{\lambda}
\end{eqnarray}

%%%%%%%%%%%%%%%%%%%%%%%%%%%%%%%%%%%%%%%%%%%%%%%%%%%%%%%%%%%%%%%%%%%%%%SUBSECTION%%%%%%%%%%%%%%%%%%%%%%%%%%%%%%%%%%%%%%%%%%%%%%%%%%%%%%%%%%%%%%%%%%%%%%
%%%%%%%%%%%%%%%%%%%%%%%%%%%%%%%%%%%%%%%%%%%%%%%%%%%%%%%%%%%%%%%%%%%%%%SUBSECTION%%%%%%%%%%%%%%%%%%%%%%%%%%%%%%%%%%%%%%%%%%%%%%%%%%%%%%%%%%%%%%%%%%%%%%
%%%%%%%%%%%%%%%%%%%%%%%%%%%%%%%%%%%%%%%%%%%%%%%%%%%%%%%%%%%%%%%%%%%%%%SUBSECTION%%%%%%%%%%%%%%%%%%%%%%%%%%%%%%%%%%%%%%%%%%%%%%%%%%%%%%%%%%%%%%%%%%%%%%
\subsection{Preliminary approach of unimodular considered}\label{subsec2}
%%%%%%%%%%%%%%%%%%%%%%%%%%%%%%%%%%%%%%%%%%%%%%%%%%%%%%%%%%%%%%%%%%%%%%SUBSECTION%%%%%%%%%%%%%%%%%%%%%%%%%%%%%%%%%%%%%%%%%%%%%%%%%%%%%%%%%%%%%%%%%%%%%%
%%%%%%%%%%%%%%%%%%%%%%%%%%%%%%%%%%%%%%%%%%%%%%%%%%%%%%%%%%%%%%%%%%%%%%SUBSECTION%%%%%%%%%%%%%%%%%%%%%%%%%%%%%%%%%%%%%%%%%%%%%%%%%%%%%%%%%%%%%%%%%%%%%%
%%%%%%%%%%%%%%%%%%%%%%%%%%%%%%%%%%%%%%%%%%%%%%%%%%%%%%%%%%%%%%%%%%%%%%SUBSECTION%%%%%%%%%%%%%%%%%%%%%%%%%%%%%%%%%%%%%%%%%%%%%%%%%%%%%%%%%%%%%%%%%%%%%%
One considers an approach to unimodular $f(T)$ gravity of setting the value of the determinant of tetrad
\begin{eqnarray}
\label{determinantfixe}
 e\equiv\det[e^{a}\,_{\mu}]=1.
\end{eqnarray}
Note $e=\sqrt{-g}$, $g$ with the determinant of the metric tensor.\par
The flat metric Friedmann-Robertson-Walker (FRW) is best suited to describe the current universe on large scale. 
The line element of the flat FRW universe can be writing 
\begin{eqnarray}
\label{universplat}
 ds^{2}=dt^{2}-a(t)^{2}(dx^2+dy^2+dz^2),
\end{eqnarray}
the metric (\ref{universplat}) can be characterized by 
\begin{eqnarray}
 e^{a}\,_{\mu}=diag[1,a(t),a(t),a(t)]\Leftrightarrow e=a^{3}(t).
\end{eqnarray}
Thus the determinant of the Tetrad varies so it can not satisfy the constraint (\ref{determinantfixe}).\par 
However, considering the parameterization given in \cite{Nojiri},
\begin{eqnarray}
\label{parametrisation}
 d\tau=a^{3}dt,
\end{eqnarray}
One has 
\begin{eqnarray}
 \label{metriqueunimodulaire}
 ds^2=a^{-6}(\tau)d\tau^2-a^{2}(\tau)(dx^2+dy^2+dz^2),
\end{eqnarray}
whether
\begin{eqnarray}
  e^{a}\,_{\mu}=diag[a(t)^{-3},a(t),a(t),a(t)]\Leftrightarrow e=1.
\end{eqnarray}
Thus the constraint (\ref{determinantfixe}) is realised, thereafter we will consider the form (\ref{metriqueunimodulaire}) instead of the form 
(\ref{universplat}) to describe our universe. \par
Our universe has experienced particular phases of evolution: after an inflationary phase it entered a phase dominated by ordinary matter and 
currently he lives a phase dominated by dark energy. It is known that the inflationary phase and phase dominated by dark energy can be described 
by a form of solution de Sitter which the scale factor is an exponential model \cite{Odintsov, Nassur}.\par 
So if we consider that the metric (\ref{universplat}) describes a de Sitter universe, then the scale factor is written 
\begin{eqnarray}
 \label{facteurdechellededesitter}
 a(t)=e^{H_{ds}t},
\end{eqnarray}
 with $H_{ds}$ Hubble parameter of de Sitter which is a constant. Then by integrating the relation (\ref{parameterization}), one  obtains 
\begin{eqnarray}
\label{tempscosmique}
 \tau=\frac{e^{3H_{ds}t}}{3H_{ds}}\Leftrightarrow t=\frac{\ln(3H_{ds}\tau)}{3H_{ds}},
\end{eqnarray}
thus, one has
\begin{eqnarray}
 \label{facteurdechellededesitterenfonctionduparametredutemps}
 a(\tau)=(3H_{ds}\tau)^{\frac{1}{3}},
\end{eqnarray}
the metric (\ref{metric unimodular}) becomes
\begin{eqnarray}
 \label{metriqueunimodulairededesitter}
 ds^{2}=(3H_{ds}\tau)^{-2}d\tau^{2}-(3H_{ds}\tau)^{\frac{2}{3}}(dx^2+dy^2+dz^2).
\end{eqnarray}\par 
It is also known that the phase dominated by ordinary matter is perfectly described by a power law evolution of the scale factor. 
So if one considers the following form\cite{Nojiri},
\begin{eqnarray}
\label{loidepuissance}
 a(t)=\left(\frac{t}{t_{0}}\right)^{h_0},
\end{eqnarray}
where $t_{0}$ and $H_0$ are constants. Using equation (\ref{loidepuissance}) in the equation (\ref{parametrisation}), one obtains
\begin{eqnarray}
\label{taudeloidepuissance}
 \tau=\frac{t_{0}}{3h_{0}+1}\left(\frac{t}{t_{0}}\right)^{3h_{0}+1},
\end{eqnarray}
whether
\begin{eqnarray}
\label{facteurdechelledelaloidepuissanceenfonctionduparametredutemps}
 a(\tau)=\left(\frac{(3h_{0}+1)\tau}{t_{0}}\right)^{\frac{h_{0}}{3h_{0}+1}}.
\end{eqnarray}
Thus equation (\ref{metriqueunimodulaire}) becomes
\begin{eqnarray}
\label{metriqueunimodulairedelaloidepuissance}
 ds^2=\left(\frac{(3h_{0}+1)\tau}{t_{0}}\right)^{-\frac{6h_{0}}{3h_{0}+1}}d\tau^{2}-\left(\frac{(3h_{0}+1)\tau}{t_{0}}\right)^{\frac{2h_{0}}{3h_{0}+1}}(dx^2+dy^2+dz^2)
\end{eqnarray}\par 
Knowing that solutions (\ref{facteurdechellededesitterenfonctionduparametredutemps}) and (\ref{facteurdechelledelaloidepuissanceenfonctionduparametredutemps})
can describe the main evolutionary stages of the universe, it would be to unify these two solutions to get a single solution that would describe
the history of the universe. It is noteworthy that for $t_{0}\longrightarrow\infty$ and $h_{0}\longrightarrow\infty$ and setting $\frac{h_{0}}{t_{0}}\equiv H_{ds}$, one obtains
the scale factor (\ref{facteurdechellededesitterenfonctionduparametredutemps}). So in these circumstances metrics (\ref{metriqueunimodulairededesitter}) and (\ref{metriqueunimodulairedelaloidepuissance})
can be linked. In other words, a solution of unification may be considered for both metrics. Especially if we consider the solution\cite{Nojiri} 
\begin{eqnarray}
 \label{unifiacation}
 ds^2=\left(\frac{\tau}{\tau_{0}}\right)^{-6f_{0}}d\tau^{2}-\left(\frac{\tau}{\tau_{0}}\right)^{2f_{0}}(dx^2+dy^2+dz^2),
\end{eqnarray}
with
\begin{eqnarray}
f_{0}\equiv\frac{h_{0}}{3h_{0}+1},\quad \tau_{0}\equiv\frac{t_{0}}{3h_{0}+1}.
\end{eqnarray}
%%%%%%%%%%%%%%%%%%%%%%%%%%%%%%%%%%%%%%%%%%%%%%%%%%%%%%%%%%%%%%%%%%%%%%SUBSECTION%%%%%%%%%%%%%%%%%%%%%%%%%%%%%%%%%%%%%%%%%%%%%%%%%%%%%%%%%%%%%%%%%%%%%%
%%%%%%%%%%%%%%%%%%%%%%%%%%%%%%%%%%%%%%%%%%%%%%%%%%%%%%%%%%%%%%%%%%%%%%SUBSECTION%%%%%%%%%%%%%%%%%%%%%%%%%%%%%%%%%%%%%%%%%%%%%%%%%%%%%%%%%%%%%%%%%%%%%%
%%%%%%%%%%%%%%%%%%%%%%%%%%%%%%%%%%%%%%%%%%%%%%%%%%%%%%%%%%%%%%%%%%%%%%SUBSECTION%%%%%%%%%%%%%%%%%%%%%%%%%%%%%%%%%%%%%%%%%%%%%%%%%%%%%%%%%%%%%%%%%%%%%%
\subsection{Unimodular $f(T)$ gravity Action and General reconstructions methods}\label{subsec3}
%%%%%%%%%%%%%%%%%%%%%%%%%%%%%%%%%%%%%%%%%%%%%%%%%%%%%%%%%%%%%%%%%%%%%%SUBSECTION%%%%%%%%%%%%%%%%%%%%%%%%%%%%%%%%%%%%%%%%%%%%%%%%%%%%%%%%%%%%%%%%%%%%%%
%%%%%%%%%%%%%%%%%%%%%%%%%%%%%%%%%%%%%%%%%%%%%%%%%%%%%%%%%%%%%%%%%%%%%%SUBSECTION%%%%%%%%%%%%%%%%%%%%%%%%%%%%%%%%%%%%%%%%%%%%%%%%%%%%%%%%%%%%%%%%%%%%%%
%%%%%%%%%%%%%%%%%%%%%%%%%%%%%%%%%%%%%%%%%%%%%%%%%%%%%%%%%%%%%%%%%%%%%%SUBSECTION%%%%%%%%%%%%%%%%%%%%%%%%%%%%%%%%%%%%%%%%%%%%%%%%%%%%%%%%%%%%%%%%%%%%%%
We will consider the action of the unimodular $f(T)$  theory  as \cite{Nojiri},
\begin{eqnarray}
 \int d^{4}x[e(f(T)-\lambda)+\lambda]+S_{matter}.
\end{eqnarray}
$S_{matter}$ the matter action, $\lambda$ the Lagrange multiplier and we asked $16\pi G=1$.
The variation of this action with respect to the tetrad leads to
\begin{eqnarray}
\label{equationdemouvement}
 S_{a}\,^{\mu\nu}\partial_{\mu}Tf''(T)-e_{a}\,^{\lambda}T^{\rho}\,_{\mu\lambda}S_{\rho}\,^{\nu\mu}f'(T)+e^{-1}\partial_{\mu}(eS_{a}\,^{\mu\nu})f'(T)+\frac{1}{4}e_{a}\,^{\nu}(f(T)-\lambda)-e_{a}\,^{\rho}\mathcal{T}_{\rho}\,^{\nu}=0,
\end{eqnarray}
where $f'(T)\equiv\frac{df(T)}{dT}$ and $f''(T)\equiv\frac{d^{2}f(T)}{dT^{2}}$.\par  
The use of relation(\ref{metriqueunimodulaire}) of the FRW metric leads to the following non-zero components of tensors,
\begin{eqnarray}
\label{tenseurdetorsion}
 T^{1}\,_{01}=T^{2}\,_{02}=T^{3}\,_{03}=\mathcal{H},\\
 K^{10}\,_{1}=K^{20}\,_{2}=K^{30}\,_{3}=-a^{6}(\tau)\mathcal{H},\\
 \label{tenseurdesuperpotentiel}
 S_{1}\,^{01}=S_{2}\,^{02}=S_{3}\,^{03}=-a^{6}(\tau)\mathcal{H}.
\end{eqnarray}
Combining relations (\ref{tenseurdetorsion}) and (\ref{tenseurdesuperpotentiel}), one obtains
\begin{eqnarray}
\label{torsionscalaire}
 T=-6a(\tau)^{6}\mathcal{H}^{2}\Rightarrow\dot{T}=-12a^{6}\mathcal{H}(3\mathcal{H}+\dot{\mathcal{H}})\Leftrightarrow-4a^{6}(3\mathcal{H}^{2}+\dot{\mathcal{H}})=\frac{\dot{T}}{3\mathcal{H}},
\end{eqnarray}
where the ``dot'' denotes the operator $\frac{d}{d\tau}$ and $\mathcal{H}\equiv\frac{1}{a(\tau)}\frac{d}{d\tau}(a(\tau))$.
Equation (\ref {torsionscalaire}) defines scalar torsion and its derivative depending on the time parameter $\tau$.\par
Considering that the energy-momentum tensor as
\begin{eqnarray}
\label{tenseurenergieimpulsion}
 \mathcal{T}_{\rho}\,^{\nu}=diag[\rho,-p,-p,-p].
\end{eqnarray}
Then, using the relations (\ref{metriqueunimodulaire}), (\ref{tenseurdetorsion}), (\ref{tenseurdesuperpotentiel}), (\ref{torsionscalaire}) and (\ref{tenseurenergieimpulsion}) in the fields equation (\ref{equationdemouvement}), one has
\begin{eqnarray}
\label{premiereequationdefriedmann0}
 -12a^{6}\mathcal{H}^{2}f'(T)+(f(T)-\lambda)-4\rho=0,\\
 \label{secondeequationdefriedmann0}
 48a^{12}\mathcal{H}^{2}(3\mathcal{H}^{2}+\dot{\mathcal{H}})f''(T)-4a^{6}(6\mathcal{H}^{2}+\dot{\mathcal{H}})f'(T)+(f(T)-\lambda)+4p=0.
\end{eqnarray}
Or equivalenty
\begin{eqnarray}
\label{premiereequationdefriedmann}
 -12a^{6}\mathcal{H}^{2}f'(T)+(f(T)-\lambda)-4\rho=0,\\
 \label{secondeequationdefriedmann}
 -4a^{6}\mathcal{H}\dot{T}f''(T)+\frac{\dot{T}}{3\mathcal{H}}f'(T)-12a^{6}\mathcal{H}^{2}f'(T)+(f(T)-\lambda)+4p=0.
\end{eqnarray}
Both versions of the modified Friedmann equations have special features for the reconstruction of the Lagrangian density
Geometric $f(T)$ on unimodular $f(T)$ gravity. The first version \Big(equations (\ref{premiereequationdefriedmann0}) and (\ref{secondeequationdefriedmann0})\Big)
 makes it easy to determine the expression of $f(T)$ in a vacuum without the knowledge of a form of scale factor.  And it leads to a Lagrange multiplier 
 $\lambda$ constant. As for the second version of the Friedmann equations \Big(equations (\ref{premiereequationdefriedmann}) and (\ref{secondeequationdefriedmann})\Big), it allows to easily determine the expression
of $f(T)$ in giving the scale factor. And it leads to a Lagrange multiplier variable, even in a vacuum.\par 
Note that for the second version of the Friedmann equations, we require the change of the scalar torsion (ie $dT\neq0)$ for
reconstruction of the function $f(T)$. 
%%%%%%%%%%%%%%%%%%%%%%%%%%%%%%%%%%%%%%%%%%%%%%%%%%%%%%%%%%%%%%%%%%%%%%Première Version%%%%%%%%%%%%%%%%%%%%%%%%%%%%%%%%%%%%%%%%%%%%%%%%%%%%%%%%%%%%%%%%%%%%%%
%%%%%%%%%%%%%%%%%%%%%%%%%%%%%%%%%%%%%%%%%%%%%%%%%%%%%%%%%%%%%%%%%%%%%%Première Version%%%%%%%%%%%%%%%%%%%%%%%%%%%%%%%%%%%%%%%%%%%%%%%%%%%%%%%%%%%%%%%%%%%%%%
%%%%%%%%%%%%%%%%%%%%%%%%%%%%%%%%%%%%%%%%%%%%%%%%%%%%%%%%%%%%%%%%%%%%%%Première Version%%%%%%%%%%%%%%%%%%%%%%%%%%%%%%%%%%%%%%%%%%%%%%%%%%%%%%%%%%%%%%%%%%%%%%
\paragraph{First case: determination of a general Lagrangian density $f(T)$}
%%%%%%%%%%%%%%%%%%%%%%%%%%%%%%%%%%%%%%%%%%%%%%%%%%%%%%%%%%%%%%%%%%%%%%Première Version%%%%%%%%%%%%%%%%%%%%%%%%%%%%%%%%%%%%%%%%%%%%%%%%%%%%%%%%%%%%%%%%%%%%%%
%%%%%%%%%%%%%%%%%%%%%%%%%%%%%%%%%%%%%%%%%%%%%%%%%%%%%%%%%%%%%%%%%%%%%%Première Version%%%%%%%%%%%%%%%%%%%%%%%%%%%%%%%%%%%%%%%%%%%%%%%%%%%%%%%%%%%%%%%%%%%%%%
%%%%%%%%%%%%%%%%%%%%%%%%%%%%%%%%%%%%%%%%%%%%%%%%%%%%%%%%%%%%%%%%%%%%%%Première Version%%%%%%%%%%%%%%%%%%%%%%%%%%%%%%%%%%%%%%%%%%%%%%%%%%%%%%%%%%%%%%%%%%%%%%
To do this, we will consider the first version of the modified Friedmann equations.
 Combining equations (\ref{premiereequationdefriedmann}) and (\ref{secondeequationdefriedmann}) in order to eliminate the term $(f(T)-\lambda)$, 
one has 
\begin{eqnarray}
\label{equationmaitraisse0}
 -2T\frac{d^{2}f(T)}{dT^{2}}-\frac{df(T)}{dT}+\frac{p+\rho}{a^{6}(3\mathcal{H}^{2}+\dot{\mathcal{H}})}=0.
\end{eqnarray}
For $\rho=0=p$, the integral of the solution of equation (\ref{equationmaitraisse0}) is given by 
\begin{eqnarray}
\label{solutionduvide}
 f(T)=-2k_{1}\sqrt{-T}+k_{2},
\end{eqnarray}
with $k_{1},k_{2}$ arbitrary constants of integration. And from equation (\ref{premiereequationdefriedmann}), it follows   
\begin{eqnarray}
 \label{multiplicateur0}
 \lambda=k_{2}.
\end{eqnarray}
Thus, the Lagrange multiplier $\lambda$ is a constant of integration in vacuum to a general approach to reconstruction
Action $f(T)$ in unimodular $f(T)$ gravity. Which joined to a philosophy of classical unimodular gravity \cite{Finkelstein}. 
%%%%%%%%%%%%%%%%%%%%%%%%%%%%%%%%%%%%%%%%%%%%%%%%%%%%%%%%%%%%%%%%%%%%%%Seconde Version%%%%%%%%%%%%%%%%%%%%%%%%%%%%%%%%%%%%%%%%%%%%%%%%%%%%%%%%%%%%%%%%%%%%%%
%%%%%%%%%%%%%%%%%%%%%%%%%%%%%%%%%%%%%%%%%%%%%%%%%%%%%%%%%%%%%%%%%%%%%%Seconde Version%%%%%%%%%%%%%%%%%%%%%%%%%%%%%%%%%%%%%%%%%%%%%%%%%%%%%%%%%%%%%%%%%%%%%%
%%%%%%%%%%%%%%%%%%%%%%%%%%%%%%%%%%%%%%%%%%%%%%%%%%%%%%%%%%%%%%%%%%%%%%Seconde Version%%%%%%%%%%%%%%%%%%%%%%%%%%%%%%%%%%%%%%%%%%%%%%%%%%%%%%%%%%%%%%%%%%%%%%
\paragraph{Second case: determination of the specific Lagrangian density}
%%%%%%%%%%%%%%%%%%%%%%%%%%%%%%%%%%%%%%%%%%%%%%%%%%%%%%%%%%%%%%%%%%%%%%Seconde Version%%%%%%%%%%%%%%%%%%%%%%%%%%%%%%%%%%%%%%%%%%%%%%%%%%%%%%%%%%%%%%%%%%%%%%
%%%%%%%%%%%%%%%%%%%%%%%%%%%%%%%%%%%%%%%%%%%%%%%%%%%%%%%%%%%%%%%%%%%%%%Seconde Version%%%%%%%%%%%%%%%%%%%%%%%%%%%%%%%%%%%%%%%%%%%%%%%%%%%%%%%%%%%%%%%%%%%%%%
%%%%%%%%%%%%%%%%%%%%%%%%%%%%%%%%%%%%%%%%%%%%%%%%%%%%%%%%%%%%%%%%%%%%%%Seconde Version%%%%%%%%%%%%%%%%%%%%%%%%%%%%%%%%%%%%%%%%%%%%%%%%%%%%%%%%%%%%%%%%%%%%%%
To do this, we will use the second version of the Friedmann equations.
Combining equations (\ref{premiereequationdefriedmann}) and (\ref{secondeequationdefriedmann}) in order to eliminate the term $(f(T)-\lambda)$, 
one obtains
\begin{eqnarray}
\label{equationmaitraisse}
 2T\dot{f}'(T)+\dot{f}(T)+4(\rho+p)=0.
\end{eqnarray}
%%%%%%%%%%%%%%%%%%%%%%%%%%%%%%%%%%%%%%%%%%%%%%%%%%%%%%%%%%%%%%%%%%%%%%SECTION%%%%%%%%%%%%%%%%%%%%%%%%%%%%%%%%%%%%%%%%%%%%%%%%%%%%%%%%%%%%%%%%%%%%%%
%%%%%%%%%%%%%%%%%%%%%%%%%%%%%%%%%%%%%%%%%%%%%%%%%%%%%%%%%%%%%%%%%%%%%%SECTION%%%%%%%%%%%%%%%%%%%%%%%%%%%%%%%%%%%%%%%%%%%%%%%%%%%%%%%%%%%%%%%%%%%%%%
%%%%%%%%%%%%%%%%%%%%%%%%%%%%%%%%%%%%%%%%%%%%%%%%%%%%%%%%%%%%%%%%%%%%%%SECTION%%%%%%%%%%%%%%%%%%%%%%%%%%%%%%%%%%%%%%%%%%%%%%%%%%%%%%%%%%%%%%%%%%%%%%
\section{Application to the specific reconstruction method in unimodular $f(T)$ gravity}\label{sec3}
%%%%%%%%%%%%%%%%%%%%%%%%%%%%%%%%%%%%%%%%%%%%%%%%%%%%%%%%%%%%%%%%%%%%%%SECTION%%%%%%%%%%%%%%%%%%%%%%%%%%%%%%%%%%%%%%%%%%%%%%%%%%%%%%%%%%%%%%%%%%%%%%
%%%%%%%%%%%%%%%%%%%%%%%%%%%%%%%%%%%%%%%%%%%%%%%%%%%%%%%%%%%%%%%%%%%%%%SECTION%%%%%%%%%%%%%%%%%%%%%%%%%%%%%%%%%%%%%%%%%%%%%%%%%%%%%%%%%%%%%%%%%%%%%%
%%%%%%%%%%%%%%%%%%%%%%%%%%%%%%%%%%%%%%%%%%%%%%%%%%%%%%%%%%%%%%%%%%%%%%SECTION%%%%%%%%%%%%%%%%%%%%%%%%%%%%%%%%%%%%%%%%%%%%%%%%%%%%%%%%%%%%%%%%%%%%%%
In what follows, we will consider the second version of the Friedmann equations (\ref{premiereequationdefriedmann}) and (\ref{secondeequationdefriedmann}).
Considering the metric (\ref{metriqueunimodulairedelaloidepuissance}), where $a(\tau)=\left(\frac{\tau}{\tau_{0}}\right)^{f_{0}}$ and
$\mathcal{H}=\frac{f_{0}}{\tau}$, it leads 
\begin{eqnarray}
\label{torsionscalairedetude}
 T=-6\left(\frac{f_{0}}{\tau_{0}}\right)^{2}\left(\frac{\tau}{\tau_{0}}\right)^{f_{0}-2}.
\end{eqnarray}
We take the contents of the universe as consisting of a barotropic fluid with the equation of state $p=\omega\rho$, then the solution of conservation equation 
$\nabla^{\mu}\mathcal{T}_{\nu\mu}=0\Leftrightarrow\dot{\rho}+3\mathcal{H}(1+\omega)\rho=0$ is geven by  
\begin{eqnarray}
 \rho=\rho_{0}a^{-3(1+\omega)}=\rho_{0}\left(\frac{\tau}{\tau_{0}}\right)^{-3(1+\omega)f_{0}},
\end{eqnarray}
for $\omega$ constant and  $\rho_{0}$ an integration constant. So the equation (\ref{equationmaitraisse}) becomes
\begin{eqnarray}
  \label{equationmaitraisse1}
\frac{d^{2}f(\tau)}{d\tau^{2}}+\frac{1}{2\tau}\frac{df(\tau)}{\tau}-\frac{(2-f_{0})\rho_{0}(1+\omega)\tau^{-3(1+\omega)f_{0}-1}}{2\tau_{0}^{-3(1+\omega)f_{0}}}=0,
\end{eqnarray}
where we have used the continuity of the derivative and the opperateur $$\frac{d}{dT}\equiv-\frac{\tau_{0}^{3}}{6f_{0}^{2}(f_{0}-2)}\left(\frac{\tau}{\tau_{0}}\right)^{-(f_{0}-3)}\frac{d}{d\tau}$$.
 Thus in vacuum ($\rho=0=p$), th equation (\ref{equationmaitraisse1}) has the solution 
\begin{eqnarray}
\label{solution1}
 f'(\tau)=\frac{C_{1}}{\sqrt{\tau}}.
\end{eqnarray}
whether
\begin{eqnarray}
 f(\tau)=2C_{1}\sqrt{\tau}+C_{2}.
\end{eqnarray}
 Using the equation (\ref{premiereequationdefriedmann}), on obtains
\begin{eqnarray}
\label{multiplicateur1}
 \lambda(\tau)%&=&2\frac{\tau_{0}}{f_{0}-2}\left(\frac{\tau}{\tau_{0}}\right)^{5f_{0}+1}f'(\tau)+f(\tau).\nonumber\\
 &=&\frac{2C_{1}\tau^{\frac{10f_{0}+1}{2}}}{(f_{0}-2)\tau_{0}^{5f_{0}}}+2C_{1}\sqrt{\tau}+C_{2}.\nonumber\\
\end{eqnarray}
where $C_{1}$ and $C_{2}$ constants of integration. Thus
\begin{eqnarray}
 f(T)=2C_{1}\sqrt{\tau_{0}}\left(\left(\frac{\tau_{0}}{f_{0}}\right)^{2}\left(-\frac{T}{6}\right)\right)^{-\frac{1}{2(2-f_{0})}}+C_{2}.
\end{eqnarray}\par
The complete solution of equation (\ref{equationmaitraisse1}) homogeneous solution which coincides with the vacuum solution (\ref{solution1}) is 
\begin{eqnarray}
 \label{solution2}
 f'(\tau)=\frac{C_{1}}{\sqrt{\tau}}-\frac{\rho_{0}(1+\omega)(2-f_{0})}{\left(6(1+\omega)f_{0}-1\right)}\left(\frac{\tau}{\tau_{0}}\right)^{-3(1+\omega)f_{0}}.
\end{eqnarray}
Whether
\begin{eqnarray}
 f(\tau)=2C_{1}\sqrt{\tau}+\frac{\rho_{0}(1+\omega)(2-f_{0})\tau_{0}}{\left(3(1+\omega)f_{0}-1\right)\left(6(1+\omega)f_{0}-1\right)}\left(\frac{\tau}{\tau_{0}}\right)^{-3(1+\omega)f_{0}+1}+C_{2}.
\end{eqnarray}
Thus
\begin{eqnarray}
\label{multiplicateur2}
 \lambda(\tau)&=&2\frac{\tau_{0}}{f_{0}-2}\left(\frac{\tau}{\tau_{0}}\right)^{5f_{0}+1}f'(\tau)+f(\tau).\nonumber\\
 &=&2\frac{\tau_{0}}{f_{0}-2}\left(\frac{\tau}{\tau_{0}}\right)^{5f_{0}+1}f'(\tau)+f(\tau).\nonumber\\
\end{eqnarray}
Therefore
\begin{eqnarray}
 f(T)&=&\frac{\rho_{0}(1+\omega)(2-f_{0})\tau_{0}}{\left(3(1+\omega)f_{0}-1\right)\left(6(1+\omega)f_{0}-1\right)}\left(\left(\frac{\tau_{0}}{f_{0}}\right)^{2}\left(-\frac{T}{6}\right)\right)^{\frac{3(1+\omega)f_{0}-1}{2-f_{0}}}\nonumber\\
 &&+2C_{1}\sqrt{\tau_{0}}\left(\left(\frac{\tau_{0}}{f_{0}}\right)^{2}\left(-\frac{T}{6}\right)\right)^{-\frac{1}{2(2-f_{0})}}+C_{2}.
\end{eqnarray}\par
Another example studied in \cite{Nojiri}
\begin{eqnarray}
 H(t)=H_{0}-\frac{M^{2}(t-t_{i})}{6}.
\end{eqnarray}
Considering the case of low cosmic time  $t$, the scale factor may be given by \cite{Nojiri}
\begin{eqnarray}
 a(t)=a_{0}e^{\frac{1}{12}\left(2(6H_{0}+M^{2}t_{i})t-(12H_{0}+M^{2}t_{i})t_{i}\right)}.
\end{eqnarray}
Where unlike Nojiri et al. we kept the term $t_{i}^{2}$ to ensure a variable scalar torsion.
 Then from the relation (\ref{parametrisation}), we show that 
\begin{eqnarray}
 \tau(t)-\tau_{0}=\frac{2a_{0}^{3}e^{\frac{1}{2}(6H_{0}+M^{2}t_{i})t}}{(6H_{0}+M^{2}t_{i})e^{\frac{1}{4}(12H_{0}+M^{2}t_{i})}}.
\end{eqnarray}
Whether
\begin{eqnarray}
 t(\tau)=\frac{\frac{(12H_{0}+M^{2}t_{i})t_{i}}{2}+2\ln\Big(\frac{(6H_{0}+M^{2}t_{i})(\tau-\tau_{0})}{2a_{0}^{3}}\Big)}{6H_{0}+M^{2}t_{i}}
\end{eqnarray}
Thus
\begin{eqnarray}
 a(\tau)=\Big(\frac{(6H_{0}+M^{2}t_{i})(\tau-\tau_{0})}{2a_{0}^{\frac{143}{48}}}\Big)^{48}.
\end{eqnarray}
And 
\begin{eqnarray}
 \mathcal{H}=\frac{48}{\tau-\tau_{0}}.
\end{eqnarray}
Whether
\begin{eqnarray}
 T=-13824\left(\frac{6H_{0}+M^{2}t_{i}}{2a_{0}^{\frac{143}{48}}}\right)^{288}(\tau-\tau_{0})^{286}.
\end{eqnarray}
The equation (\ref{equationmaitraisse}) becomes 
\begin{eqnarray}
\label{equationmaitraisse2}
 \frac{d^{2}f(\tau)}{d\tau^{2}}+\frac{1}{2(\tau-\tau_{0})}\frac{df(\tau)}{d\tau}+286\rho_{0}\left(\frac{6H_{0}+M^{2}t_{i}}{2a_{0}^{\frac{143}{48}}}\right)^{-144(1+\omega)}(\tau-\tau_{0})^{-144(1+\omega)-1}=0.
\end{eqnarray}
where the continuity of the derivative was used, and the operator $$\frac{d}{dT}-\frac{1}{3953664}\left(\frac{6H_{0}+M^{2}t_{i}}{2a_{0}^{\frac{143}{48}}}\right)^{-288}(\tau-\tau_{0})^{-285}\frac{d}{d\tau}$$.
In vacuum,  the equation (\ref{equationmaitraisse2}) has a solution
\begin{eqnarray}
 f'(\tau)=\frac{C_{5}}{\sqrt{\tau-\tau_{0}}}.
\end{eqnarray}
Whether
\begin{eqnarray}
\label{action1}
 f(\tau)=2C_{5}\sqrt{\tau-\tau_{0}}+C_{6}.
\end{eqnarray}
thus 
\begin{eqnarray}
\label{multiplicateur3}
 \lambda(\tau)&=&\frac{(\tau-\tau_{0})}{1144}f'(\tau)+f(\tau).\nonumber\\
 &=&\frac{2289C_{5}}{1144}\sqrt{\tau-\tau_{0}}+C_{6}.
\end{eqnarray}
Therefore the action (\ref{action1}) can be re-write as  
\begin{eqnarray}
 f(T)=2a_{0}^{3/2}C_{5}\left(\frac{2}{6H_{0}+M^{2}t_{i}}\right)^{\frac{72}{143}}\left(-\frac{T}{13824}\right)^{\frac{1}{572}}+C_{6}.
\end{eqnarray}\par 
Furthermore, the complete solution of the equation (\ref{equationmaitraisse2}) is given by 
\begin{eqnarray}
 f'(\tau)=\frac{C_{5}}{\sqrt{\tau-\tau_{0}}}+\left(\frac{572\rho_{0}}{288\omega+287}\right)\left(\left(\frac{6H_{0}+M^{2}t_{i}}{2a_{0}^{\frac{143}{48}}}\right)(\tau-\tau_{0})\right)^{-144(1+\omega)}.
\end{eqnarray}
Whether
\begin{eqnarray}
 f(\tau)=2C_{5}\sqrt{\tau-\tau_{0}}-\frac{572\rho_{0}}{(144\omega+143)(288\omega+287)}\left(\frac{6H_{0}+M^{2}t_{i}}{2a_{0}^{\frac{143}{48}}}\right)^{-144(1+\omega)}(\tau-\tau_{0})^{-144(1+\omega)+1}+C_{6}.
\end{eqnarray}
From the equation (\ref{premiereequationdefriedmann}), one obtains
\begin{eqnarray}
\label{multiplicateur4}
 \lambda(\tau)&=&\frac{2289C_{5}\sqrt{\tau-\tau_{0}}}{1144}+\frac{\rho_{0}(72\omega+71)}{(144\omega+143)(288\omega+287)}\left(\frac{6H_{0}+M^{2}t_{i}}{2a_{0}^{\frac{143}{48}}}\right)^{-144(1+\omega)}(\tau-\tau_{0})^{-144(1+\omega)+1}\nonumber\\
 &&-4\rho_{0}\Big(\frac{(6H_{0}+M^{2}t_{i})(\tau-\tau_{0})}{2a_{0}^{\frac{143}{48}}}\Big)^{-144(1+\omega)}+C_{6}.
\end{eqnarray}
Furthermore
\begin{eqnarray}
 f(T)&=&-\frac{572\rho_{0}}{(144\omega+143)(288\omega+287)}\left(\frac{6H_{0}+M^{2}t_{i}}{2a_{0}^{\frac{143}{48}}}\right)^{\frac{144\omega}{143}}\left(-\frac{T}{13824}\right)^{-\frac{144(1+\omega)-1}{286}}\nonumber\\
 &&+2C_{5}\left(\frac{6H_{0}+M^{2}t_{i}}{2a_{0}^{\frac{143}{48}}}\right)^{-\frac{72}{143}}\left(-\frac{T}{13824}\right)^{\frac{1}{572}}+C_{6}.
\end{eqnarray}
thus we see that the second version of the Friedmann equations leads to the Lagrange multipliers are be functions of time parameter $\tau$ even in a 
vacuum (ie $\rho=0$) \Big( see equations (\ref{multiplicateur1}), (\ref{multiplicateur2}), (\ref{multiplicateur3}) and (\ref{multiplicateur4})\Big).
Unlike the first version of the Friedmann equations that leaves invariant the Lagrange multiplier (\ref{multiplicateur0}) in a vacuum.
%%%%%%%%%%%%%%%%%%%%%%%%%%%%%%%%%%%%%%%%%%%%%%%%%%%%%%%%%%%%%%%%%%%%%%SECTION%%%%%%%%%%%%%%%%%%%%%%%%%%%%%%%%%%%%%%%%%%%%%%%%%%%%%%%%%%%%%%%%%%%%%%
%%%%%%%%%%%%%%%%%%%%%%%%%%%%%%%%%%%%%%%%%%%%%%%%%%%%%%%%%%%%%%%%%%%%%%SECTION%%%%%%%%%%%%%%%%%%%%%%%%%%%%%%%%%%%%%%%%%%%%%%%%%%%%%%%%%%%%%%%%%%%%%%
%%%%%%%%%%%%%%%%%%%%%%%%%%%%%%%%%%%%%%%%%%%%%%%%%%%%%%%%%%%%%%%%%%%%%%SECTION%%%%%%%%%%%%%%%%%%%%%%%%%%%%%%%%%%%%%%%%%%%%%%%%%%%%%%%%%%%%%%%%%%%%%%
\section{Conclusion}\label{sec4}
%%%%%%%%%%%%%%%%%%%%%%%%%%%%%%%%%%%%%%%%%%%%%%%%%%%%%%%%%%%%%%%%%%%%%%SECTION%%%%%%%%%%%%%%%%%%%%%%%%%%%%%%%%%%%%%%%%%%%%%%%%%%%%%%%%%%%%%%%%%%%%%%
%%%%%%%%%%%%%%%%%%%%%%%%%%%%%%%%%%%%%%%%%%%%%%%%%%%%%%%%%%%%%%%%%%%%%%SECTION%%%%%%%%%%%%%%%%%%%%%%%%%%%%%%%%%%%%%%%%%%%%%%%%%%%%%%%%%%%%%%%%%%%%%%
%%%%%%%%%%%%%%%%%%%%%%%%%%%%%%%%%%%%%%%%%%%%%%%%%%%%%%%%%%%%%%%%%%%%%%SECTION%%%%%%%%%%%%%%%%%%%%%%%%%%%%%%%%%%%%%%%%%%%%%%%%%%%%%%%%%%%%%%%%%%%%%%
This work consists of reconstructing  $f(T)$ actions with the constraint of the tetrad determinant fixed to the unit, in unimodulaire $f(T)$ gravity.
Then after the different general approaches related to embrace unimodular $f(T)$ gravity, we started our work by establishing two equivalent 
versions of the modified Friedmann equations.\par
These two equivalent versions of Friedmann equations have provided us specific $f(T)$ functions reconstruction approaches.
One that allows the reconstruction of a function $f(T)$ General which is independent of any expression of the scale factor $a(t)$ and
requires the Lagrange multiplier being equal to an integration constant $(\lambda\equiv constant)$, in vacuum. And the other which allows reconstruction
special actions $f(T)$, by giving the expression of the scale factor $a(t)$ and resulting in expression of the time-dependent Lagrange 
multipliers $(\lambda\equiv\lambda(\tau))$ even in a vacuum.

%%%%%%%%%%%%%%%%%%%%%%%%%%%%%%%%%%%%%%%%%%%%%%%%%%%%%%%%%%%%%%%%%%%%%%%%%%%%%%%%%%%%%%%%%%%%%%%%%%%%%%%%%%%%%%%%%%%%%%%%%
%%%%%%%%%%%%%%%%%%%%%%%%%%%%%%%%%%%%%%%%%%%%%%%%%%%%%%%%%%%%%%%%%%%%%%%%%%%%%%%%%%%%%%%%%%%%%%%%%%%%%%%%%%%%%%%%%%%%%%%%%

\vspace{0.5cm}
{\bf Acknowledgement}: S. B. Nassur thanks DAAD for financial support. 
%%%%%%%%%%%%%%%%%%%%%%%%%%%%%%%%%%%%%%%%%%%%%%%%%%%%%%%%%%%%%%%%%%%%%%%%%%%%%%%%%%%%%%%%%%%%%%%%%%%%%%%%%%%%%%%%%%%%%%%%%
%%%%%%%%%%%%%%%%%%%%%%%%%%%%%%%%%%%%%%%%%%%%%%%%%%%%%%%%%%%%%%%%%%%%%%%%%%%%%%%%%%%%%%%%%%%%%%%%%%%%%%%%%%%%%%%%%%%%%%%%%
%%%%%%%%%%%%%%%%%%%%%%%%%%%%%%%%%%%%%%%%%%%%%%%%%%%%%%%%%%%%%%%%%%%%%%%%%%%%%%%%%%%%%%%%%%%%%%%%%%%%%%%%%%%%%%%%%%%%%%%%%

%On a posé $\kappa^{2}=8\pi G=1$ et
\end{document}